# Composite stacks for reliable > 17 T trapped fields in bulk superconductor magnets

Kai Yuan Huang[1], Yunhua Shi[1], Jan Srpčič[1], Mark D Ainslie[1], Devendra K Namburi[1], Anthony R Dennis[1], Difan Zhou[2], Martin Boll[3], Mykhaylo Filipenko[3], Jan Jaroszynski[4], Eric E Hellstrom[4], David A Cardwell[1] and John H Durrell[1]

[1] Department of Engineering, University of Cambridge, Cambridge CB2 1PZ, United Kingdom

[2] Department of Physics, Shanghai University, Shanghai 200444, China

[3] Siemens AG Corporate Technology eAircraft, Willy-Messerschmitt-Str. 1, D-82024, Taufkirchen, Germany

[4] Applied Superconductivity Center, National High Magnetic Field Laboratory, Florida State University, 2031 East Paul Dirac Drive, Tallahassee, FL 32310, United States of America

Email: kyh30@cam.ac.uk

## Abstract

Trapped fields of over 20 T are, in principle, achievable in bulk, single-grain high temperature cuprate superconductors. The principle barriers to realizing such performance are, firstly, the large tensile stresses that develop during the magnetization of such trapped-field magnets as a result of the Lorentz force, which lead to brittle fracture of these ceramic-like materials at high fields and, secondly, catastrophic thermal instabilities as a result of flux movement during magnetization. Moreover, for a batch of samples nominally fabricated identically, the statistical nature of the failure mechanism means the best performance (i.e. trapped fields of over 17 T) cannot be attained reliably. The magnetization process, particularly to higher fields, also often damages the samples such that they cannot repeatedly trap high fields following subsequent magnetization. In this study, we report the sequential trapping of magnetic fields of ~ 17 T, achieving 16.8 T at 26 K initially and 17.6 T at 22.5 K subsequently, in a stack of two Ag-doped $GdBa_2Cu_3O_{7-\delta}$ bulk superconductor composites of diameter 24 mm reinforced with (1) stainless-steel laminations, and (2) shrink-fit stainless steel rings. A trapped field of 17.6 T is, in fact, comparable with the highest trapped fields reported to date for bulk superconducting magnets of any mechanical and chemical composition, and this was achieved using the first composite stack to be fabricated by this technique. These post-melt-processing treatments, which are relatively straightforward to implement, were used to improve both the mechanical properties and the thermal stability of the resultant composite structure, providing what we believe is a promising route to achieving reliably fields of over 20 T.

## 1. Introduction

The (RE)Ba$_2$Cu$_3$O$_{7-\delta}$ family of bulk high temperature superconductors [or (RE)BCO; where RE = rare earth element or yttrium] can be used as trapped-field magnets [1–3] with superior performance compared with conventional hard ferromagnets. Trapped fields of just over 17 T are achievable by bulk superconductors, which is around an order of magnitude greater than the fields attainable by the best conventional permanent magnets [1]. Most notably, Tomita and Murakami demonstrated in 2003 a trapped field of 17.24 T at 29 K in a stack of two YBCO samples of diameter 26.5 mm reinforced by resin and alloy impregnation along with carbon fiber wrapping [4]. In 2014, Durrell *et al.* exceeded this performance by demonstrating a trapped field of 17.6 T at 26 K in a stack of two 24 mm diameter GdBCO-Ag bulk superconductors reinforced with shrink-fit stainless steel rings [5]. There has also been an increasing amount of interest in using stacks of high temperature superconducting (HTS) tape as composite bulks, with Patel *et al.* reporting a trapped field of 17.66 T at 8 K in a 34.4 mm diameter stack in 2018 [6].

It should be possible to achieve trapped fields of over 20 T in (RE)BCO bulk superconductors with the current state-of-the-art material processing techniques that enable the fabrication of large, well-connected superconducting grains with excellent superconducting properties [i.e. critical current density $J_c(B, T)$] [7]. However, two performance-limiting factors need to be addressed for these fields to be realized. Firstly, bulk superconductors exhibit fairly low tensile strength due to the large number of inherent small-scale defects, such as pores and micro-cracks, which are produced during the melt-process. Since the Lorentz force, $\boldsymbol{F_L} = \boldsymbol{J_c} \times \boldsymbol{B}$, leads to large magnetic stresses during the magnetization of such trapped field magnets [8, 9], these brittle materials suffer frequently from mechanical failure (i.e. cracking). Secondly, (RE)BCO materials generally exhibit relatively low thermal conductivity [10, 11], and so, when a significant amount of heat, $Q = \boldsymbol{E} \cdot \boldsymbol{J}$, is generated due to flux movement during magnetization, catastrophic thermal instabilities and flux jumps can occur [4, 12].

To address these issues and to achieve record trapped fields of over 17 T, bulk superconducting samples have been reinforced by various techniques, including resin and alloy impregnation, carbon fiber wrapping [4] and shrink-fit stainless steel rings [5]. It is clear, therefore, that adequate mechanical reinforcement and thermal stability are key to achieving a trapped field greater than 20 T, which, in turn, could strengthen the case for practical applications such as compact nuclear magnetic resonance (NMR) and magnetic resonance imaging (MRI) systems, magnetic separation and magnetically targeted drug delivery [13–15].

In this study, we report the sequential high-field field-cooled magnetization of a stack of two Ag-doped GdBa$_2$Cu$_3$O$_{7-\delta}$ bulk superconductor composites of diameter 24 mm, which trapped fields of 16.8 T at 26 K initially and subsequently 17.6 T at 22.5 K. A trapped field of 17.6 T is, in fact, comparable with the highest trapped fields reported to date for bulk superconducting magnets of any mechanical and chemical composition. To form the composite assembly, stainless steel discs were sandwiched between layers of bulk superconductor to define a strengthened laminated structure. These results are also particularly significant since the trapped fields reported here were achieved using the first composite stack to be fabricated by this technique. This is in stark contrast to the magnetization of standard, as-grown bulk superconductors where, for a batch of samples fabricated identically, the statistical nature of the failure mechanism means the best performance (i.e. trapped fields of over 17 T) cannot be attained reliably. A typical failure behavior of 'standard' bulk magnets was also demonstrated by Tomita and Murakami [4].

The reinforcement concept was, in fact, first proposed by Morita *et al.* in 2017 for ring-shaped bulk magnets [16, 17], where they demonstrated that the proposed structure significantly decreased the mechanical strains experienced by the superconductor and prevented crack formation. Furthermore, the resultant composite proposed here also benefits from improved thermal stability as a result of enhanced thermal capacitance and thermal conductivity, which helps to reduce the occurrence of flux jumps.

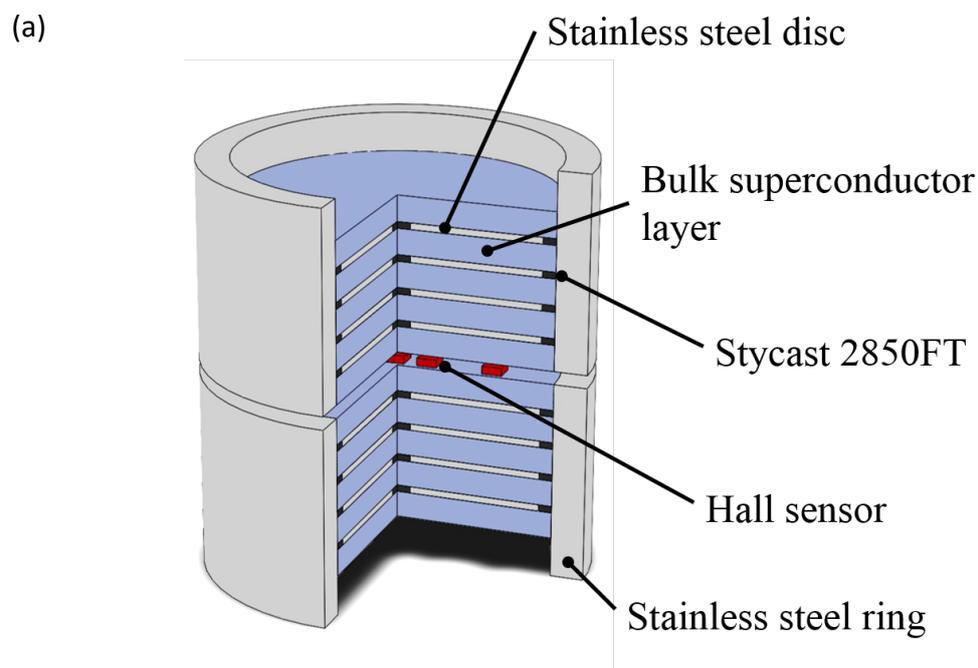

(a) Stainless steel disc / Bulk superconductor layer / Stycast 2850FT / Hall sensor / Stainless steel ring

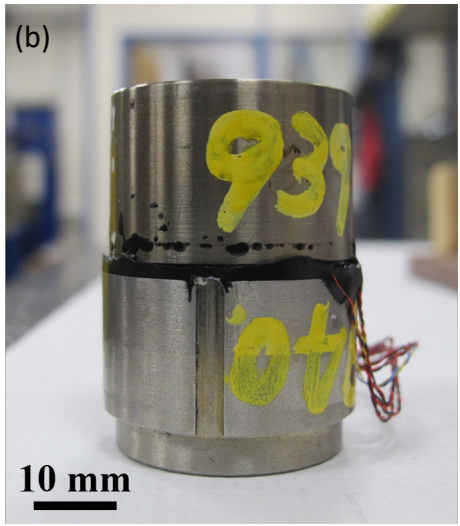
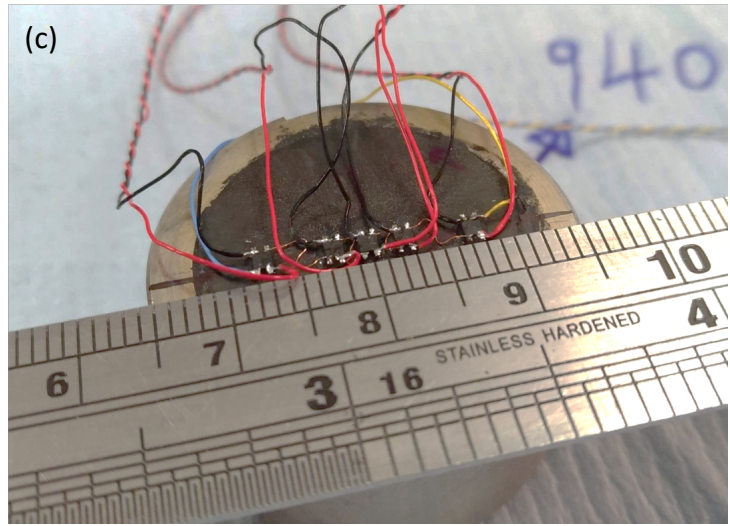

Figure 1. (a) Partial section view of a schematic illustration of the bulk superconductor composite structure implemented in this study, showing the layers of bulk superconductor and stainless steel. The magnetic field was measured with Hall sensors placed in the center of the stack. (b) Photograph of the two-sample stack. (c) Positions of the five Hall sensors mounted in-between the two samples.

## 2. Numerical simulations

### 2.1 Field-trapping potential

To design and assess appropriate reinforcement arrangements, two-dimensional (2D) axisymmetric finite-element models, implemented in the commercial finite element software package *COMSOL Multiphysics*, were used to study the field-trapping potential and mechanical stability of different bulk superconductor structures during the field-cooled magnetization process. Further details on the modeling arrangement and the implementation of the $J_c(B, T)$ data can be found at [18–20]. In summary, the 'Magnetic Field Formulation', 'Heat Transfer in Solids' and 'Solid Mechanics' interfaces were coupled together to allow for a comprehensive study of the thermal stresses $\sigma_\theta^{COOL}$ arising from differential thermal contraction and the electromagnetic stresses $\sigma_\theta^{FCM}$ arising from interaction between the current and the magnetic field. The $J_c(B, T)$ characteristics used in the numerical simulations were measured for a representative GdBCO-Ag bulk specimen for fields of up to 6 T over a temperature range of 30 – 92 K and extended to 20 T using the equation proposed by Jirsa *et al.* [21], as described in [18]. A list of the assumed relevant material properties used in the models, including references to the relevant literature, can be found in Appendix and/or ref. [18].

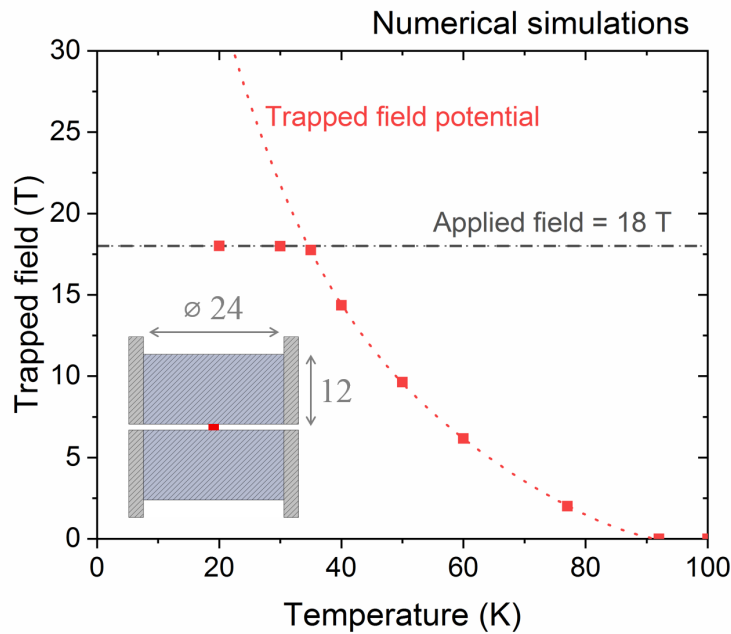

Figure 2. Numerical simulation results for the trapped magnetic field achievable at the center of a conventional two-sample stack at various temperatures with an applied field of 18 T, indicated by the solid red squares. In addition, the field potentially achievable in the absence of mechanical limitations and with a sufficiently high applied field is shown by the red dotted line. The inset shows the sample geometry used in the numerical simulations. The applied

field and sample dimensions chosen are very similar to those reported in other experimental studies [4, 5, 22].

The trapped field achievable theoretically in a conventional stack of bulk superconductors was determined taking into account only $J_c(B, T)$ and ignoring mechanical limitations of the sample. Figure 2 shows the temperature-dependence of the theoretical trapped field at the center of the stack. With an applied field of 18 T and at temperatures below 35 K, the trapped field is no longer limited by the current-carrying capability of the superconductor, but rather by the magnitude of the applied field. This is also in strong agreement with the trapped field profiles measured by Durrell *et al.* and Tomita and Murakami at 26 K and 29 K, respectively [4, 5], where both reports observed a flattened field distribution towards the center of the bulk stack, suggesting that the bulk superconductors were not fully saturated and that higher trapped fields were feasible at these temperatures.

Figure 2 implies that trapped fields of over 20 T and 30 T are achievable at 30 K and 20 K, respectively, and that the volume fraction of superconducting material in a composite structure could be potentially reduced and still trap a high trapped field at a reasonable temperature, since the $J_c$ of a state-of-the-art bulk superconductor is sufficiently large at lower temperatures.

2.2 Composite bulk vs. standard bulk

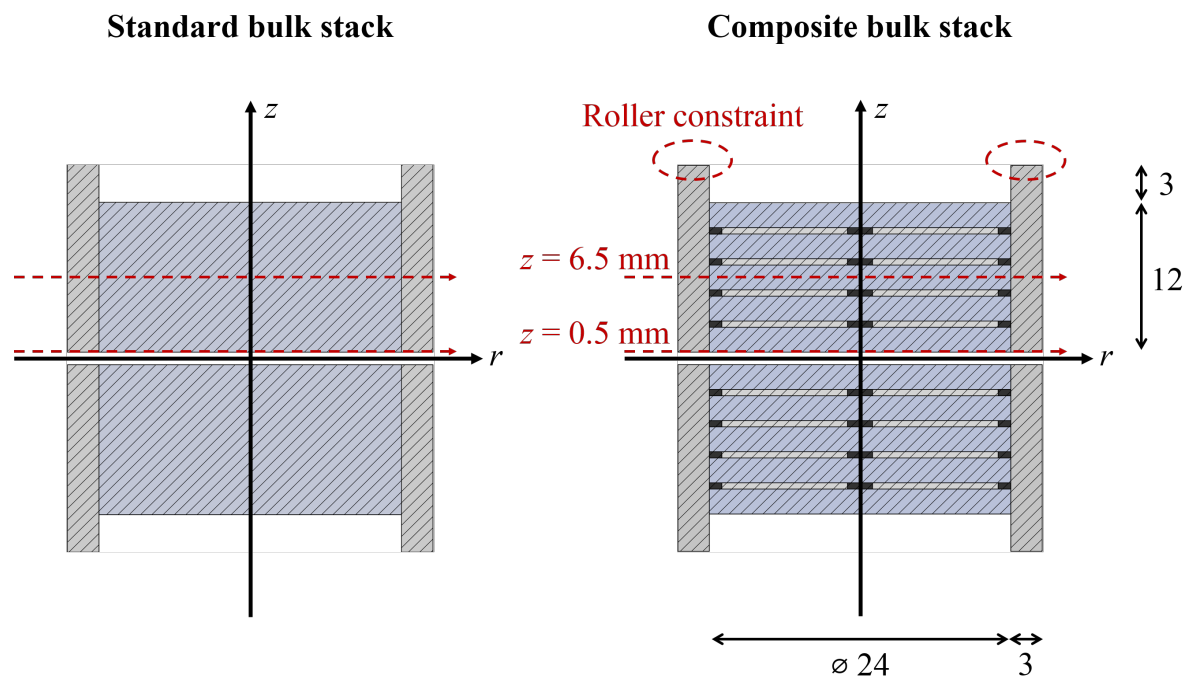

Figure 3. Cross-sectional views of the 'standard' conventional stack and composite stack used in the numerical simulations. The lines marked $z = 0.5$ mm and $z = 6.5$ mm correspond,

respectively, to the top surface and mid-plane of the bulk superconductors in the upper half of each structure.

Figure 1(a) shows a schematic illustration of the composite bulk structure used in this study. This is very similar to the laminated structure proposed by Morita *et al.*; however, their approach utilized silver sputter deposition, further heat treatment and joining of the superconductor layers via solder [16]. Figure 3 shows a cross-sectional view of the geometries used in the numerical simulations. A stack without the stainless-steel laminations, referred to here as the 'standard' stack, was also modeled for comparison. Positions $z$ = 0.5 mm and $z$ = 6.5 mm have been marked in the figure, corresponding to the surface and mid-plane of the upper bulk sample in each case. There is continuity of displacements across every interface, which is the default boundary condition built into the COMSOL 'Solid Mechanics' module. This implies perfect mechanical connection between adjacent materials at every interface and that the stresses will be transferred accordingly. A 'roller' constraint was applied to the top of the stainless-steel rings, as shown in figure 3, to simulate attaching the bulk stack to the end of a measurement probe. Finally, only the upper halves of the stacks (i.e. $z > 0$) were modeled in view of the symmetry of the arrangement.

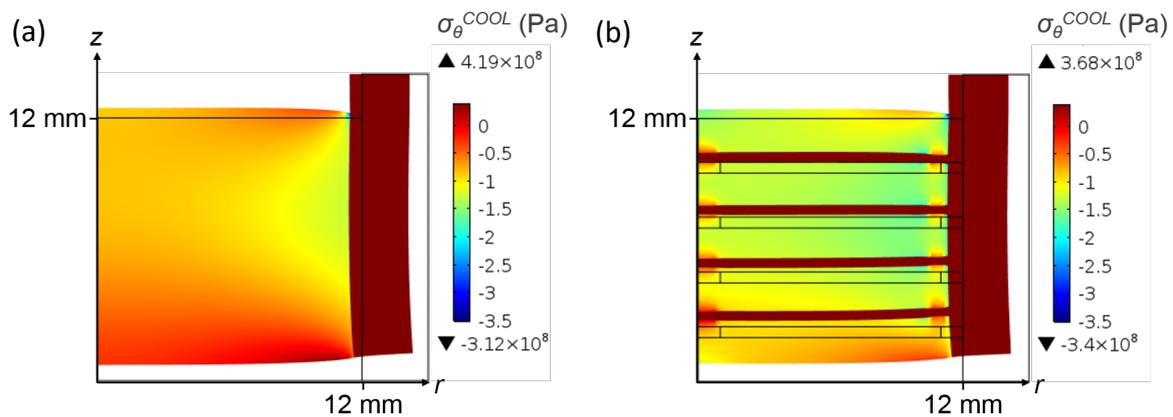

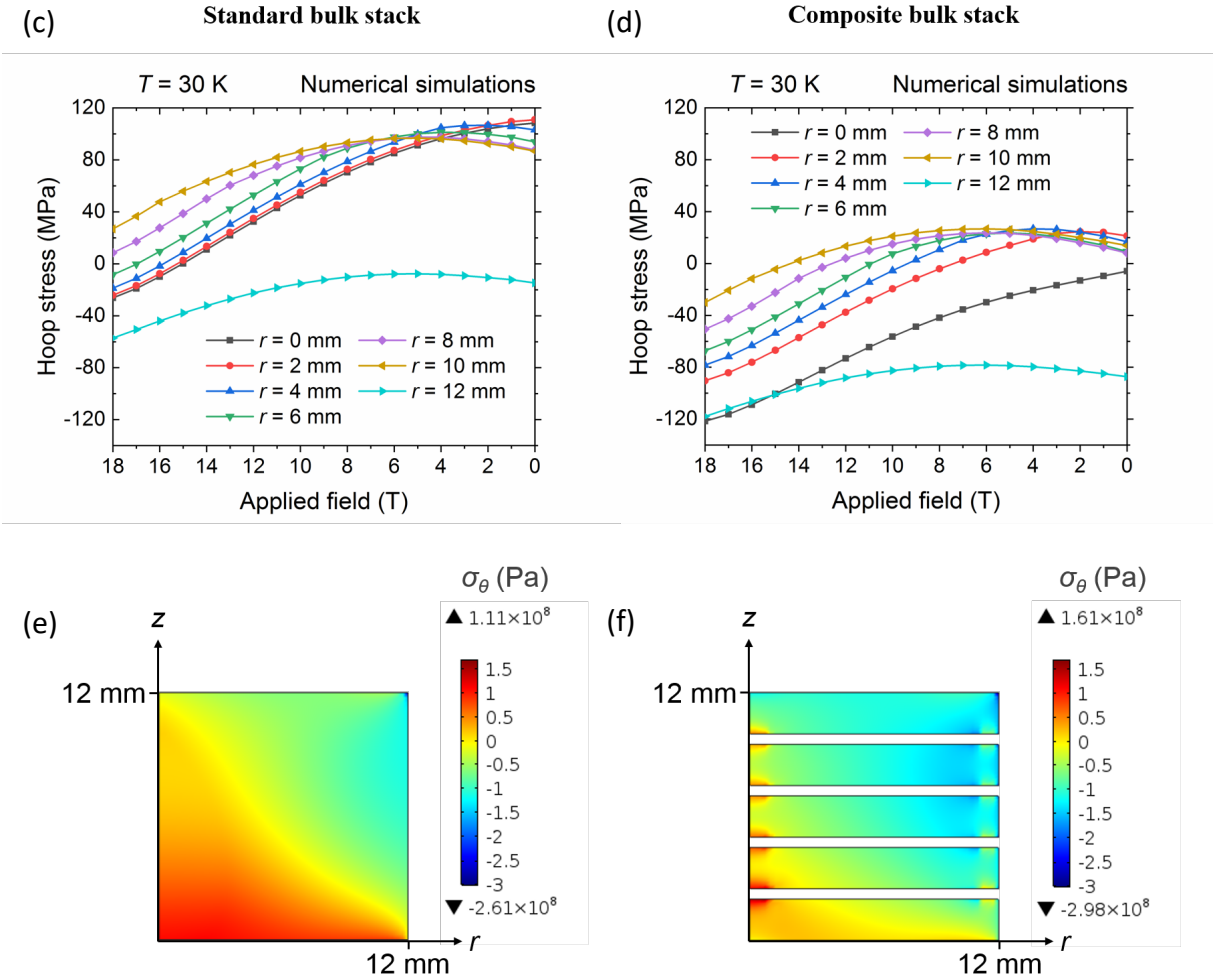

Figure 4. (a) Thermal hoop stress $\sigma_\theta^{COOL}$ throughout the cross-section of the standard stack at 30 K before the magnetization process. The deformation (exaggerated by a scale of 25) is also shown here to demonstrate the effect of differential thermal contraction along the $z$-direction. (b) Thermal hoop stress $\sigma_\theta^{COOL}$ throughout the cross-section of the composite stack at 30 K before the magnetization process. (c) Hoop stress experienced by the standard stack along $z$ = 0.5 mm during the ramp down of the applied field from 18 T to 0 T at 30 K. (d) Hoop stress experienced by the composite stack along $z$ = 0.5 mm during the ramp down of the applied field from 18 T to 0 T at 30 K. (e) Hoop stress throughout the cross-section of the standard stack at the end of the magnetization process. The maximum and minimum stresses are shown. (f) Hoop stress throughout the cross-section of the composite stack (showing only the superconducting layers) at the end of the magnetization process. Again, the maximum and minimum stresses are shown.

Figure 4(a) shows the thermal hoop stress $\sigma_\theta^{COOL}$ throughout the cross-section of the standard stack just before the field was ramped down. The deformation has been exaggerated by × 25 for the standard stack in order to illustrate the way it deforms upon cooling. The stainless steel ring contracts more than the bulk superconductor along the $z$- direction when cooled to 30 K due to the difference in thermal expansion coefficients, causing the upper and lower surfaces of the bulk superconductor to be somewhat bowed and in tension. On the other hand, figure 4(b) shows the thermal hoop stress in the composite stack at the same point in time and plotted using the same color scale, for which it can be seen that the inclusion of the stainless-steel laminations results in a more compressive stress state prior to ramping down the field. This is due to the stainless steel discs contracting more quickly than the bulk superconductor along the *a-b* direction, causing the bulk layers to be significantly compressed by the neighboring epoxy/steel layers.

Figures 4(c) and 4(d) show the evolution of the hoop stresses along the *z* = 0.5 mm plane in both stacks as the field was ramped down from 18 T at 30 K. The stresses shown at 18 T (i.e. at the beginning of the magnetization process) correspond solely to the thermal hoop stresses $\sigma_\theta^{COOL}$ shown in figures 4(a) and 4(b). Changes in the hoop stress as the field was ramped down correspond to the electromagnetic contribution $\sigma_\theta^{FCM}$.

It can be seen the composite arrangement experienced a thermal hoop stress that is almost 100 MPa more compressive at the center and around 60 MPa more compressive towards the edge of the bulk superconductor. The total hoop stress in the standard stack reached as high as 111 MPa during the magnetization process, whereas the maximum hoop stress in the composite stack was only 27 MPa. It is clear the standard stack is unlikely to survive the magnetization process based on the flexural strength of ~ 105 MPa and splitting tensile strength of ~ 34 MPa measured by Konstantopoulou *et al.* for GdBCO-Ag at 77 K [23]. Furthermore, due to the increased stiffness of the composite structure, the electromagnetic hoop stress $\sigma_\theta^{FCM}$ was also suppressed by around 20 MPa when compared to the standard stack. It is clear from figures 4(e) and 4(f) that the volume of superconductor under high tensile stress has been reduced significantly with the incorporation of the stainless-steel discs, with most of the high tensile stresses confined to the stainless steel layers, as shown in figure 4(b).

The numerical simulations also revealed that the peak trapped field at the center of the composite stack is not affected by the inclusion of the stainless-steel layers, with both the standard stack and composite stack trapping 18 T at the end of the magnetization process. However, it must be noted the trapped field profile for the composite stack was sharper in comparison to the standard stack (i.e. the field decays more quickly with radial position).

Furthermore, positive impacts that are difficult to demonstrate with the simulation include the ability of the composite to resist catastrophic fracture across the whole bulk, since the stainless steel sheets (with high fracture toughness and high tensile strength) prevent cracks from propagating from one bulk layer to another, the possibility of filling the exposed pores and defects of each superconductor layer once they have been cut from the parent bulk superconductor, which can increase significantly the tensile strength of the superconductor itself [24], and, finally, by essentially replacing superconducting material with stainless steel, the fractional volume of heat source producing $Q = \boldsymbol{E} \cdot \boldsymbol{J}$ is reduced significantly, reducing the probability of flux jumps and improving the thermal stability of the composite.

### 3. Experimental details

3.1 Sample fabrication and mechanical reinforcement

Single-grain, Ag-doped $GdBa_2Cu_3O_{7-\delta}$ bulk superconductor samples (GdBCO-Ag), ~ 25 mm in diameter and ~ 10 mm in thickness, were fabricated by top seeded melt growth, as described in detail elsewhere [25]. To ensure they were single grains, each sample was magnetized with an applied field of 1.5 T at 77 K and the trapped field profile was mapped. Both samples exhibited a conical trapped field geometry, and a peak trapped field of 0.99 T and 1.01 T.

To form the laminated composite shown in figure 1, the bulk superconductors were cut using a diamond saw into five slices of approximately equal thickness. Four discs of stainless steel (grade 304), 22 mm in diameter and 0.51 mm in thickness, were then sandwiched in-between the GdBCO-Ag slices, and glued together using Stycast® 2850 FT (mixed with 23 LV catalyst). A disc thickness of 0.51 mm was chosen (equivalent to 24 AWG) since the cutting process removed around 0.5 mm of bulk material per slice. The composites were then machined down to a diameter of 24.1 mm, before a stainless steel ring (also grade 304), with an inner diameter of 24.0 mm and outer diameter of 29.0 mm, was heated to 300 °C prior to shrink-fitting over each bulk composite arrangement.

A linear array of five Lakeshore HGT-2101 Hall sensors was placed in-between the stack at radial positions of –7.5, –2.5, 0, 2.5 and 7.5 mm from the center of the stack, as shown in figure 1(c). The two composite bulk samples were then combined into a stack using Stycast epoxy resin with their top surfaces (i.e. the position of the seed crystals during melt-processing) both pointing towards the center of the stack.

3.2 Trapped field measurements

The output voltage of each Hall sensor was measured as a function of external applied field in order to determine the field measured at discrete points at the center of the stack. The Hall sensors were driven by a 22 Hz, 10 µA peak sine wave generated by a Keithley 6221 current source, and the Hall voltage was measured using a lock-in amplifier for each sensor. A calibrated Cernox sensor on the end of the stack was used to measure the temperature of the stack, while the sample temperature was controlled and stabilized using a wire-wound heater wrapped around the stack.

The stacks were magnetized in the bore of the 18 T SCM2 system at the National High Magnetic Field Laboratory (NHMFL), Florida State University. The field-cooled magnetization procedure was as follows: an external field of 18 T was applied whilst the stack temperature was held at 100 K (i.e. above critical temperature $T_c$). The stack was cooled to the desired temperature and allowed to stabilize once the field was stable at 18 T. Finally, the external field was removed at a rate of 0.015 – 0.02 T/min, and the resultant trapped field and subsequent flux creep were measured at the temperature set-point.

## 4. Results and discussion

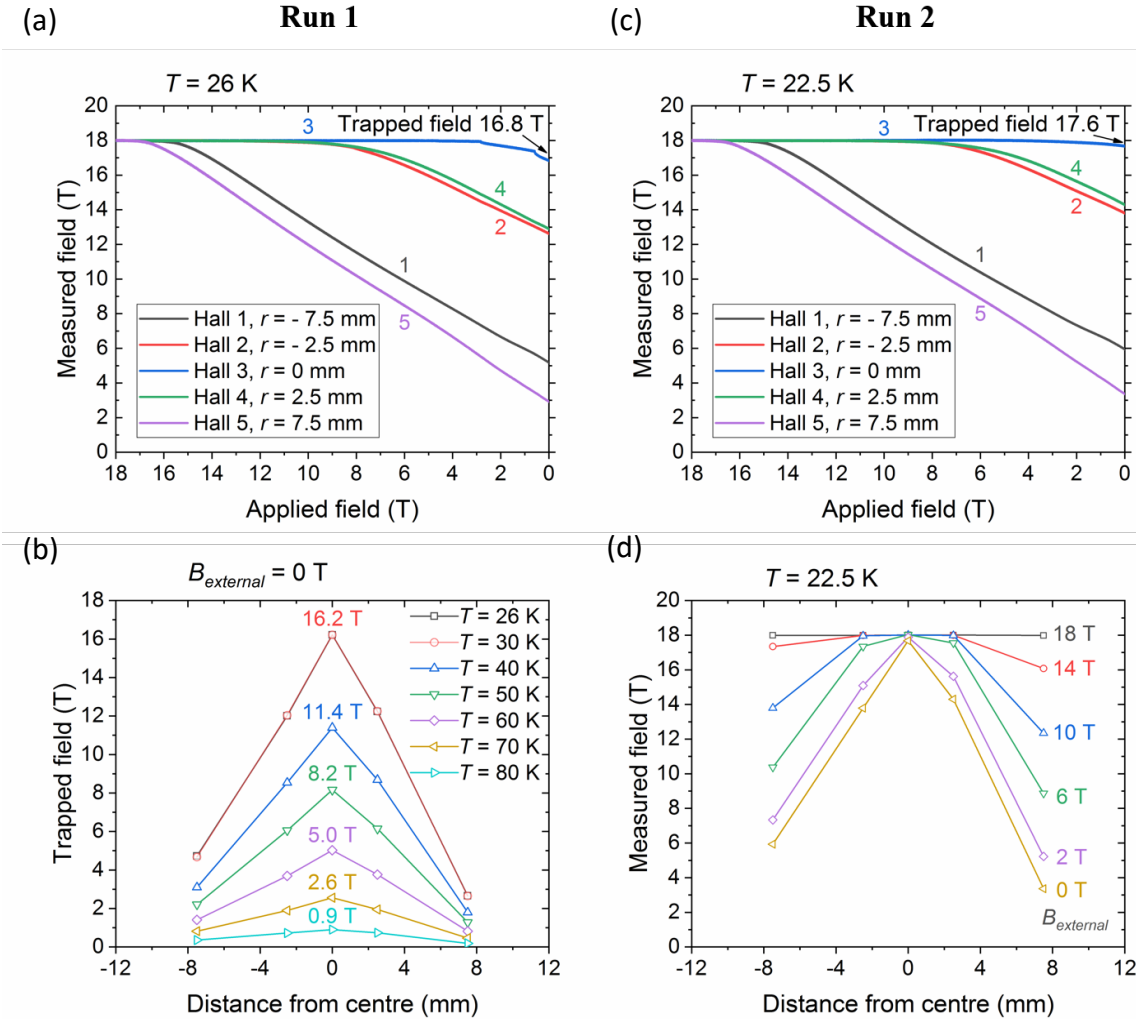

Figure 5. (a) Magnitude of the trapped field measured at the center of the two-sample composite stack at 26 K as the external applied field was ramped-down. (b) Trapped field profiles measured at various temperatures as the sample stack was warmed slowly at a rate of approximately 1 K/min. The field measured by the central Hall sensor is shown at 10 K temperature increments. (c) Magnitude of the field measured at the center of the two-sample composite stack at 22.5 K as the external applied field was ramped-down. (d) Field profiles measured at various points during the ramp-down of the applied field as the stack temperature was maintained at 22.5 K.

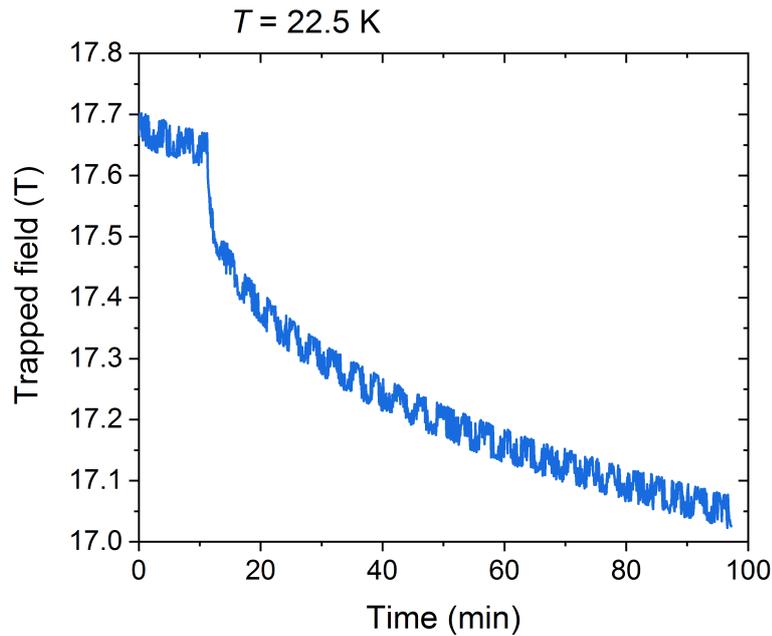

Figure 6. Decay of the trapped field as a result of flux creep immediately after the complete removal of the applied field. The stack temperature was maintained at 22.5 K throughout the measurement.

The composite stack was magnetized initially by field cooling in 18 T at 26 K with a constant ramp-down rate of 0.02 T/min, which resulted in a trapped field of 16.8 T immediately after the complete removal of the applied field. The fields measured at five positions across the center of the stack during the magnetization process are shown in figure 5(a). The sample was then warmed slowly at a rate of approximately 1 K/min to determine the temperature dependence of the trapped field distribution, as shown in figure 5(b).

Minor changes in the gradient of the field at $r$ = 0 mm towards the end of the magnetization process, evident in figure 5(a), hint at the additional mechanical reinforcement and thermal stability provided by this composite structure. It is possible that this behavior corresponds to the beginning of crack formation or flux jumps that could have led to an avalanche within the bulk superconductor layers, although they appear to be suppressed by the composite structure with enhanced mechanical and thermal stability. For instance, if a crack has formed in one of the bulk superconductor layers, crack propagation may have been prevented by the neighboring stainless steel layers, obviating catastrophic damage. Similarly, if regions of the bulk superconductor layers begin to heat-up due to flux movement, the heat could have been dissipated quickly by the local stainless steel.

The stack was re-magnetized using a similar process to confirm reproducibility, since it did not show any evidence of catastrophic failure during the initial magnetization process. This involved field cooling the stack in a field of 18 T at 22.5 K following a constant ramp-down rate of 0.015 T/min, which produced a trapped field of just over 17.6 T immediately after the complete removal of the applied field. The trapped field was averaged over 100 seconds after the external field reached zero to minimize errors due to signal noise. The fields measured during the magnetization are shown in figure 5(c). Slight asymmetry towards the edge of the stack can be seen in figure 5(d), with the fields measured by the Hall sensors positioned at $r = \pm 7.5$ mm differing by 2.5 T, which may be due to minor asymmetry in the $J_c(B)$ properties of one or both of the composite bulk superconductors, a small error in the positioning of the Hall sensors, or to a small misalignment of the two bulk samples.

Subsequently, to determine the decay in trapped field with time as a result of flux creep, the trapped field was recorded for 100 minutes with the sample temperature held at 22.5 K, as shown in figure 6. The flux creep measured in this composite stack appears to be more severe than that reported by Durrell *et al.* in their 17.6 T stack of two GdBCO bulk superconductors [5].

The effectiveness of the proposed composite, as predicted by detailed numerical simulations presented in Section 2, has been verified by the high trapped field achieved experimentally, which is comparable to the highest trapped fields reported to date. In addition, these results are significant because reliable and reproducible magnetization with an applied field of 18 T at temperatures below 30 K without the sample failing mechanically or thermally has been shown here to be possible. Despite being the only stack of this structure that was measured, it was able to survive two high-field magnetization tests, first achieving 16.8 T at 26 K and then 17.6 T at 22.5 K. This is noteworthy because, as Durrell *et al*. pointed out in their paper, the two-sample stack that trapped 17.6 T failed when it was re-magnetized subsequently and two other stacks (produced identically to the 17.6 T sample) that were measured in the same experiment could only trap 10 T and 15.4 T [5]. This implies high variability amongst 'standard' stacks of bulk superconductors is due most likely to variation in the defect distribution from bulk to bulk. To the best of our knowledge, previous studies have not been able to successfully re-magnetize a single bulk sample or a multi-sample stack under such high field conditions.

On the whole, the composite bulk arrangement can be perceived as a hybrid between a standard bulk superconductor disc and a stack of tapes. Stacked tapes have superior mechanical and thermal stability as a result of their large metallic volume fraction (> 90 %), although they require a low operating temperature due to the low engineering current density, $J_e$. On the other hand, bulk superconductors have high $J_c$ (of the order of $10^9$ A/m$^2$ below 50 K), which means higher operating temperatures can be used, but often fail mechanically or thermally due to their low tensile strength and low thermal conductivity. The composite structure reported here can be viewed as a practical compromise between the two types of trapped field magnets. This is also evident from the temperature required to achieve trapped fields of around 17.6 T [5], which was achieved at 26 K for the bulk superconductor stack, whereas the stacked tapes had to be cooled to 8 K [6], and the composite structure reported here to 22.5 K.

## 5. Conclusions

We have demonstrated the sequential high-field magnetization of a stack of two Ag-doped $GdBa_2Cu_3O_{7-\delta}$ bulk superconductor composites of diameter 24 mm, formed by sandwiching layers of high-strength and high-stiffness stainless steel in-between layers of bulk superconductor. The aim of these post-melt-processing treatments was to increase the strength and toughness of the resultant laminate as well as improving the thermal stability of the composite. After field-cooled magnetization, the stack successfully trapped 16.8 T at 26 K initially, followed by 17.6 T at 22.5 K in a subsequent experiment. The composite stack, which was the first constructed with this reinforcement technique, was able to achieve a trapped field comparable to the highest trapped fields reported to date for bulk superconducting magnets of any mechanical and chemical composition. This is important because, as Durrell *et al*. noted in their report of a trapped field of 17.6 T, the stack that trapped the record field of 17.6 T failed when it was re-magnetized subsequently and two other stacks that were fabricated identically to that of the 17.6 T sample could only trap 10 T and 15.4 T, demonstrating the high variability amongst notionally 'standard' stacks.

Based on both simulation and experimental results, we have shown that fields of over 20 T should be attainable in bulk superconductor magnets based on state-of-the-art $J_c(B, T)$ properties and that the reinforcement technique presented in this study may prove to be a promising route to ensure these trapped field magnets are sufficiently stable, mechanically and thermally, to sustain practical trapped fields of over 20 T.

**Acknowledgements**

The authors would like to acknowledge support from Siemens AG Corporate Research and the UK Engineering and Physical Sciences Research Council (EPSRC, grant EP/P00962X/1 and EP/P020313/1). A portion of this work was performed at the National High Magnetic Field Laboratory, which is supported by the National Science Foundation Cooperative Agreement No. DMR-1644779 and the State of Florida. Mr Kai Yuan Huang would also like to thank Tomáš Hlásek and Jan Plecháček from CAN Superconductors for supplying a large number of bulk superconductor samples for mechanical testing. Dr Mark Ainslie would like to acknowledge financial support from an EPSRC Early Career Fellowship EP/P020313/1. Data supporting this publication are available from the University of Cambridge Institutional Repository at https://doi.org/10.17863/CAM.42982.

**Appendix**

| Parameter | Description | Value | Reference(s) |
|---|---|---|---|
| $n$ | $n$ value ($E$–$J$ power law) | 20 | [26] |
| $E_0$ | Characteristic voltage ($E$–$J$ power law) | $1 \times 10^{-4}$ V m$^{-1}$ | [26] |
| $J_c(B, T)$ | Critical current density | Interpolation | [19, 20] |
| $E_{bulk}$ | Young's modulus (bulk) | $1 \times 10^{11}$ Pa | [27–29] |
| $E_{stainless}$ | Young's modulus (stainless steel) | $1.93 \times 10^{11}$ Pa | [28, 29] |
| $E_{epoxy}$ | Young's modulus (epoxy resin) | $9 \times 10^{9}$ Pa | [30] |
| $v_{bulk}$ | Poisson's ratio (bulk) | 0.25 | [31] |
| $v_{stainless}$ | Poisson's ratio (stainless steel) | 0.28 | [28, 29] |
| $v_{epoxy}$ | Poisson's ratio (epoxy resin) | 0.33 | [32] |
| $a_{bulk}$ | Coefficient of thermal expansion (bulk) | $5.2 \times 10^{-6}$ K$^{-1}$ | [27–29] |
| $a_{stainless}$ | Coefficient of thermal expansion (stainless steel) | $1.27 \times 10^{-5}$ K$^{-1}$ | [28, 29] |
| $a_{epoxy}$ | Coefficient of thermal expansion (epoxy resin) | $3.9 \times 10^{-5}$ K$^{-1}$ | [30] |

Table 1. List of assumed material properties for the two-dimensional axisymmetric models described in Section 2.

# References

[1] A. M. Campbell and D. A. Cardwell, "Bulk high temperature superconductors for magnet applićations," *Cryogenics (Guildf).*, vol. 37, no. 10, pp. 567–575, 1997.

[2] M. Murakami, "Progress in applications of bulk high temperature superconductors," *Supercond. Sci. Technol.*, vol. 13, no. 5, p. 448, 2000.

[3] J. R. Hull and M. Murakami, "Applications of bulk high-temperature superconductors," *Proc. IEEE*, vol. 92, no. 10, pp. 1705–1718, 2004.

[4] M. Tomita and M. Murakami, "High-temperature superconductor bulk magnets that can trap magnetic fields of over 17 tesla at 29 K," *Nature*, vol. 421, no. 6922, p. 517, 2003.

[5] J. H. Durrell *et al.*, "A trapped field of 17.6 T in melt-processed, bulk Gd-Ba-Cu-O reinforced with shrink-fit steel," *Supercond. Sci. Technol.*, vol. 27, no. 8, p. 82001, 2014.

[6] A. Patel *et al.*, "A trapped field of 17.7 T in a stack of high temperature superconducting tape," *Supercond. Sci. Technol.*, vol. 31, no. 9, p. 09LT01, 2018.

[7] H. Fujishiro, T. Naito, and S. Awaji, "Proposal of effective mechanical reinforcement structure for REBaCuO disk bulk pair by full metal encapsulation to achieve higher trapped field over 20 T," *Supercond. Sci. Technol.*, 2019.

[8] Y. Ren, R. Weinstein, J. Liu, R. P. Sawh, and C. Foster, "Damage caused by magnetic pressure at high trapped field in quasi-permanent magnets composed of melt-textured Y-Ba-Cu-O superconductor," *Phys. C Supercond.*, vol. 251, no. 1–2, pp. 15–26, 1995.

[9] T. H. Johansen, "Flux-pinning-induced stress and magnetostriction in bulk superconductors," *Supercond. Sci. Technol.*, vol. 13, no. 10, p. R121, 2000.

[10] K. Noto *et al.*, "Thermal and mechanical properties of high Tc bulk superconductors and their applications," *Phys. C Supercond.*, vol. 392, pp. 677–683, 2003.

[11] H. Fujishiro, S. Nariki, and M. Murakami, "Thermal conductivity and thermoelectric power of DyBaCuO bulk superconductors," *Supercond. Sci. Technol.*, vol. 19, no. 7, p. S447, 2006.

[12] L. Gao, Y. Y. Xue, R. L. Meng, and C. W. Chu, "Thermal instability, magnetic field shielding and trapping in single-grain YBa2Cu3O7− δ bulk materials," *Appl. Phys. Lett.*, vol. 64, no. 4, pp. 520–522, 1994.

[13] N. Saho, N. Nishijima, H. Tanaka, and A. Sasaki, "Development of portable superconducting bulk magnet system," *Phys. C Supercond.*, vol. 469, no. 15–20, pp. 1286–1289, 2009.

[14] T. Nakamura *et al.*, "Development of a superconducting bulk magnet for NMR and MRI," *J. Magn. Reson.*, vol. 259, pp. 68–75, 2015.

[15] J. H. Durrell *et al.*, "Bulk superconductors: A roadmap to applications," *Supercond. Sci. Technol.*, vol. 31, no. 10, p. 103501, 2018.

[16]  K. Morita, H. Teshima, and S. Nariki, "Development of New Reinforcement Method and 10T Magnetization of QMG Magnet," *Nippon Steel Sumitomo Met. Tech. Rep.*, vol. No. 117, 2017.

[17]  M. Morita, "History and recent progress of QMG$^{TM}$ and QMG bulk magnets," in *Journal of Physics: Conference Series*, 2018, vol. 1054, no. 1, p. 12046.

[18]  M. D. Ainslie *et al.*, "Numerical modelling of mechanical stresses in bulk superconductor magnets with and without mechanical reinforcement," *Supercond. Sci. Technol.*, vol. 32, no. 3, p. 34002, 2019.

[19]  M. D. Ainslie *et al.*, "Enhanced trapped field performance of bulk high-temperature superconductors using split coil, pulsed field magnetization with an iron yoke," *Supercond. Sci. Technol.*, vol. 29, no. 7, p. 74003, 2016.

[20]  M. D. Ainslie, D. Zhou, H. Fujishiro, K. Takahashi, Y. H. Shi, and J. H. Durrell, "Flux jump-assisted pulsed field magnetisation of high-J c bulk high-temperature superconductors," *Supercond. Sci. Technol.*, vol. 29, no. 12, p. 124004, 2016.

[21]  M. Jirsa, L. Pust, D. Dlouhý, and M. R. Koblischka, "Fishtail shape in the magnetic hysteresis loop for superconductors: Interplay between different pinning mechanisms," *Phys. Rev. B*, vol. 55, no. 5, p. 3276, 1997.

[22]  G. Fuchs *et al.*, "Trapped magnetic fields larger than 14 T in bulk YBa 2 Cu 3 O 7− x," *Appl. Phys. Lett.*, vol. 76, no. 15, pp. 2107–2109, 2000.

[23]  K. Konstantopoulou, Y. H. Shi, A. R. Dennis, J. H. Durrell, J. Y. Pastor, and D. A. Cardwell, "Mechanical characterization of GdBCO/Ag and YBCO single grains fabricated by top-seeded melt growth at 77 and 300 K," *Supercond. Sci. Technol.*, vol. 27, no. 11, p. 115011, 2014.

[24]  M. Tomita and M. Murakami, "Improvement of the mechanical properties of bulk superconductors with resin impregnation," *Supercond. Sci. Technol.*, vol. 13, no. 6, p. 722, 2000.

[25]  Y. Shi *et al.*, "Batch-processed GdBCO–Ag bulk superconductors fabricated using generic seeds with high trapped fields," *Phys. C Supercond. its Appl.*, vol. 470, no. 17–18, pp. 685–688, 2010.

[26]  M. D. Ainslie and H. Fujishiro, "Modelling of bulk superconductor magnetization," *Supercond. Sci. Technol.*, vol. 28, no. 5, p. 53002, 2015.

[27]  H. Mochizuki, H. Fujishiro, T. Naito, Y. Itoh, Y. Yanagi, and T. Nakamura, "Trapped field characteristics and fracture behavior of REBaCuO bulk ring during pulsed field magnetization," *IEEE Trans. Appl. Supercond.*, vol. 26, no. 4, pp. 1–5, 2015.

[28]  H. Fujishiro *et al.*, "Simulation studies of mechanical stresses in REBaCuO superconducting ring bulks with infinite and finite height reinforced by metal ring during field-cooled magnetization," *Supercond. Sci. Technol.*, vol. 30, no. 8, p. 85008, 2017.

[29]  K. Takahashi, H. Fujishiro, T. Naito, Y. Yanagi, Y. Itoh, and T. Nakamura, "Fracture behavior analysis of EuBaCuO superconducting ring bulk reinforced by a stainless

steel ring during field-cooled magnetization," *Supercond. Sci. Technol.*, vol. 30, no. 11, p. 115006, 2017.

[30] Henkel, "LOCTITE STYCAST 2850FT." Technical Data Sheet, 2016.

[31] P. Diko, G. Fuchs, and G. Krabbes, "Influence of silver addition on cracking in melt-grown YBCO," *Phys. C Supercond.*, vol. 363, no. 1, pp. 60–66, 2001.

[32] Epoxy Technology Inc, "Understanding Mechanical Properties of Epoxies For Modeling, Finite Element Analysis (FEA)," 2012. [Online]. Available: http://www.epotek.com/site/files/Techtips/pdfs/tip19.pdf.